# Simultaneous weak value amplification of angular Goos-Hänchen and Imbert-Fedorov shifts in partial reflection

S. Goswami,[1,+] M. Pal,[1,+] A. Nandi,[1] P.K.Panigrahi,[1] N.Ghosh[1,*]

[1] Dept. of Physical Sciences, Indian Institute of Science Education and Research-Kolkata, Mohanpur Campus, Mohanpur 741 252, India

[+] *These authors contributed equally to this work*

*Corresponding author: nghosh@iiserkol.ac.in*

The optical analogue of quantum weak measurements have shown considerable promise for the amplification and observation of tiny optical beam shifts, namely, the Goos-Hänchen (GH) and the Imbert-Fedorov (IF) shifts. Here, we demonstrate simultaneous weak value amplification of both the *angular* GH and the IF shifts in *partial reflection* of fundamental Gaussian beam at planar dielectric interfaces. We employ pre- and post selection schemes with appropriate linear polarization basis states for simultaneous weak measurements and amplification of both these shifts. The experimentally observed enhancement of the beam shifts and their dependence on the angle of incidence are analyzed / interpreted via quantum mechanical treatment of weak measurements.

The reflection/transmission, or total internal reflection (TIR) of a physical light beam at planar dielectric interfaces exhibit intriguing behavior which are not governed by the laws of geometrical optics [1-4]. Unlike plane waves, finite beams exhibit longitudinal (in the plane of incidence) and transverse (perpendicular to the plane of incidence) shifts at planar interfaces. Such effects may manifest as a spatial displacement (coordinate shift) or an angular deflection (momentum shift) of the beam depending upon the nature of the interaction [3,4]. These so-called longitudinal Goos-Hänchen (GH) and the transverse Imbert-Federov (IF) shifts have evoked intensive investigations owing to fundamental interests and potential applications [3,4]. While the GH shift originates from the angular gradient of the reflection / transmission coefficients, the IF shift has its origin in spin orbit interaction of light and conservation of total angular momentum [3]. A variant of the latter effect is the so-called Spin Hall effect of light (SHEL) [5,6]. The eigen polarization modes of the GH shift are TM (p) and TE (s) linearly polarized waves and that of SHEL are left and right circularly (or elliptically) polarized waves respectively [3,7,8]. The eigenmodes of IF shift on the other hand can either be left/right circular (elliptical) polarization and +45º/-45º linear polarization [3]. These phenomena are rather tiny and accurate measurement, unique interpretation of these shifts is extremely challenging. Among the various techniques employed to observe these effects, weak measurements have been particularly promising and fundamentally rich [5, 7 -13]. The weak measurement of the beam shifts is an optical analogue of the well known quantum mechanical weak measurements [8,9]. Here, the optical beam shifts (GH, IF and SH) act as weak measuring effects and the state of polarization serves as pre and post selection mechanisms. Indeed the first observation of SHEL in transmission of light beam at an air/glass interface was reported using weak measurements [5]. Similarly, weak value amplification of the *spatial* GH shift and *spatial* IF shifts have been reported for TIR in glass/air interfaces [11, 12]. Amplification of the beam shifts using weak measurements have accordingly been modeled either via classical optics treatment [14, 15] or using quantum mechanical description [7-9]. It is pertinent to note that both the GH (for input p and s linear polarization modes) and the IF (for circular / elliptical and +45º/-45º linear polarization modes) shifts are spatial in nature in the case of TIR due to the fact that the Fresnel reflection coefficients ($r_p$ and $r_s$) are complex [3,4,7]. In contrast, the GH shift in partial (non-total) reflection (wherein $r_p$ and $r_s$ are real) is purely angular and the IF shift can either be spatial (for circular/elliptical polarization modes) or angular (for +45º/-45º linear polarization modes) [4]. Among the two variants of the GH shifts, while the spatial shift in TIR has been observed long back, experimental observation of the angular GH shift in partial reflection has only been reported recently [16].

In this letter, we report simultaneous weak value amplification of both the *angular* GH and IF shifts in *partial reflection* of fundamental Gaussian beam at planar air/glass interface. We employ pre- and post selection schemes with appropriate linear polarization basis states for simultaneous weak value amplification of the *angular* GH and the IF shifts. Coupling of these two shifts and their simultaneous weak measurement manifested as a diagonal beam shift (with respect to the plane of incidence), which could be successfully detected and quantified. We also subsequently performed weak measurements on IF shifts alone by nullifying the weak measurement on angular GH shift, using pre-selection of s or p linear polarization states. We analyze/interpret the various intriguing manifestations of the angular GH and

IF shifts in partial reflection using quantum mechanical treatment of weak measurements.

Even though the amplification of the angular GH and IF shifts for a given pre and post selection of polarization states can be described classically [14], we adopt quantum mechanical description. The latter approach yields better physical insight with relatively simpler mathematical treatment. In this approach, we define the quantum operators for the GH and IF shift for the case of partial reflection as [9]

$$GH = \begin{bmatrix} \Omega_p(\theta) & 0 \\ 0 & \Omega_s(\theta) \end{bmatrix}$$

$$IF = \begin{bmatrix} 0 & \Omega_l(\theta) \\ -\Omega_r(\theta) & 0 \end{bmatrix} \quad (1)$$

Where, $\Omega_p(\theta) = -i\frac{\partial \ln r_p}{\partial \theta}$, $\Omega_s(\theta) = -i\frac{\partial \ln r_s}{\partial \theta}$

$$\Omega_l(\theta) = i\left(1 + \frac{r_p}{r_s}\right)\cot\theta, \quad \Omega_r(\theta) = i\left(1 + \frac{r_s}{r_p}\right)\cot\theta \quad (2)$$

Here, $\theta$ is the angle of incidence and $r_p$, $r_s$ are the angle-dependent amplitude reflection coefficients for p and s polarized waves. Note that for $\theta$ very close to the Brewster angle ($\theta \to \theta_B$), the above expressions need corrections incorporating higher order terms [13,15,16].

In the above description, the state of polarization of light is represented by the two component Jones vector $|\psi\rangle = [\alpha_1, \alpha_2]^T$ (e.g., the p and s linear polarizations are represented as $[1,0]^T$ and $[0,1]^T$ respectively). The corresponding Fresnel reflection Jones matrix is

$$R = \begin{bmatrix} r_p(\theta) & 0 \\ 0 & r_s(\theta) \end{bmatrix} \quad (3)$$

In weak measurements, pre ($\psi_{pre}$) and post selections ($\psi_{post}$) of the polarization states determine the so-called weak values of the GH and IF shift matrices

$$A_w^{GH} = \frac{\langle \psi_{post}|GH|\psi_{pre}\rangle}{\langle \psi_{post}|\psi_{pre}\rangle} \text{ and } A_w^{IF} = \frac{\langle \psi_{post}|IF|\psi_{pre}\rangle}{\langle \psi_{post}|\psi_{pre}\rangle} \quad (4)$$

Note that the pre-selected state here is not actually the input polarization state ($\psi_{in}$), rather the state is ($|\psi pre\rangle = R|\psi in\rangle$). In weak measurements, the initial and the final states ($\psi_{pre}$ and $\psi_{post}$) are chosen to be nearly (but not exactly) orthogonal, which results in large magnitude of the weak value ($A_w$) leading to large amplification of the beam shifts. In what follows, we describe the specific schemes adopted by us for simultaneous weak value amplification of angular GH and IF shifts. We define the input state as $|\psi in\rangle = [\alpha_1, \alpha_2]^T$ so that the pre-selected polarization state becomes $|\psi pre\rangle = [r_p\alpha_1, r_s\alpha_2]^T$. Henceforth, we shall omit the normalization factors of Jones vectors. We employ two different pre and post selection schemes all using linear polarization basis, for weak value amplification of the shifts. The first scheme enable coupling and simultaneous weak value amplification of both the *angular* GH and the IF shifts. The second scheme decouples the two and perform weak value amplification of the IF shift alone.

**Scheme 1:** $\alpha_1 = 1, \alpha_2 = 1$

The corresponding pre and post selection states are

$$|\psi_{pre}\rangle = \frac{1}{\sqrt{2}}[r_p \quad r_s]^T$$

$$|\psi_{post}\rangle = \frac{1}{\sqrt{2}}[r_s \pm \varepsilon r_p \quad -r_p \pm \varepsilon r_s]^T \quad (5)$$

Here, $\varepsilon$ is a small angle by which the post selected state is away from the orthogonal (to the pre selection) state. The expressions for the weak values of the GH and IF shift matrices for this scheme can be derived using Eq. 4 as

$$A_w^{GH\pm} = \frac{(r_p^2\Omega_p + r_s^2\Omega_s)}{(r_p^2 + r_s^2)} \pm \frac{1}{\varepsilon}\frac{r_p r_s(\Omega_p - \Omega_s)}{(r_p^2 + r_s^2)} \quad (6a)$$

$$A_w^{IF\pm} = \frac{r_p r_s(\Omega_l - \Omega_r)}{(r_p^2 + r_s^2)} \pm \frac{1}{\varepsilon}\frac{(r_p^2\Omega_l + r_s^2\Omega_r)}{(r_p^2 + r_s^2)} \quad (6b)$$

The $\pm$ signs correspond to post selections with $\pm \varepsilon$. The dimensionless beam shift parameters (actual beam shifts are proportional to these) between the two post selected states ($\pm \varepsilon$ away from orthogonal) can thus be written as

$$\Delta_w^{GH} = \frac{2}{\varepsilon}\frac{r_p r_s(\Omega_p - \Omega_s)}{(r_p^2 + r_s^2)} = \frac{2i}{\varepsilon}\frac{(r_s r_p' - r_p r_s')}{(r_p^2 + r_s^2)} \quad (7a)$$

$$\Delta_w^{IF} = \frac{2}{\varepsilon}\frac{r_p r_s(\Omega_l + \Omega_r)}{(r_p^2 + r_s^2)} = \frac{2i}{\varepsilon}\frac{(r_s + r_p)^2}{(r_p^2 + r_s^2)}\cot\theta \quad (7b)$$

Where $r_p'$ ($r_s'$) are angular derivatives of $r_p$ ($r_s$).

As apparent from Eq. 7, both the shift parameters are imaginary (in partial reflection) and accordingly the amplified shifts are angular in nature. Thus, pre and post selection in linear polarization basis simultaneously amplifies the angular GH shift (between p and s linear polarization modes) and the IF shift (between left/right circular or elliptical polarizations). Moreover, this converts spatial IF shift to angular one [12, 14]. Note, weak value amplification of the beam shifts using this scheme leads to an additional scaling of the amplification by a factor of $2r_p r_s/(r_p^2 + r_s^2)$ (which itself depends upon $\theta$).

**Scheme 2:** $\alpha_1 = 1, \alpha_2 = 0$ (p-state) or $\alpha_1 = 0, \alpha_2 = 1$ (s-state)

$$|\psi_{pre}\rangle = [1 \quad 0]^T \text{ (p-state); } |\psi_{pre}\rangle = [0 \quad 1]^T \text{ (s-state)}$$

$$|\psi_{post}\rangle = [\pm\varepsilon \quad 1]^T \text{ (p-state); } |\psi_{post}\rangle = [1 \quad \pm\varepsilon]^T \text{ (s-state);} \quad (8)$$

For this pre selection, there would be no weak value amplification of the angular GH shift. However, the IF shift would undergo amplification and the corresponding expressions of the amplified shift parameters for pre selection with p-state and s-state (respectively) becomes

$$\Delta_{w,p}^{IF} = \frac{2}{\varepsilon}\Omega_r = \frac{2i}{\varepsilon}(1 + \frac{r_s}{r_p})\cot\theta; \quad \Delta_{w,s}^{IF} = \frac{2}{\varepsilon}\Omega_l = \frac{2i}{\varepsilon}(1 + \frac{r_p}{r_s})\cot\theta \quad (9)$$

The dimensionless imaginary shift parameters (representing angular shifts) of Eqs. 7 and 9 are related to the shift in the position of the centroid of a beam in a direction parallel ($<x>$ for GH) and perpendicular ($<y>$ for IF) to the plane of incidence as [12, 14]

$$\langle x \rangle = \frac{\lambda}{2\pi i}\frac{z}{z_o}\Delta_w^{GH}; \quad \langle y \rangle = \frac{\lambda}{2\pi i}\frac{z}{z_o}\Delta_w^{IF} \quad (10)$$

Where $\lambda$ is the wavelength of light, z is the propagation distance and $z_o$ is the Rayleigh range of a fundamental Gaussian beam.

We now turn to the experimental system (**Figure 1**) employed for realizing the weak measurement schemes. Briefly, the 632.8 nm line of a He-Ne laser (HRR120-1, Thorlabs, USA) was used to seed the system. The beam was spatially filtered, collimated and then focused by a lens L (focal length f = 20 cm) to a spot size of $\omega_0 \sim 130$ μm.

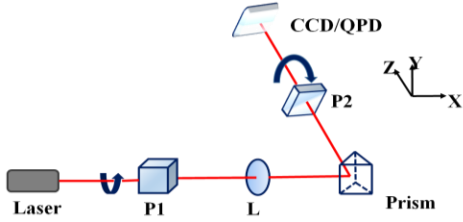

**Fig.1**. A schematic of the set-up for the weak measurement of the angular GH and IF shifts in partial reflection. $P_1$, $P_2$: rotatable Glan-Thompson linear polarizers mounted on precision rotation mount; L: Lens, QPD: Quadrant photodiode. The prism mounted on a precision rotation stage act as the weak measuring device. $P_1$ prepares the input state and $P_2$ post selects the final state.

The polarization of the incident beam is controlled by a rotatable Glan-Thompson linear polarizer $P_1$ (GTH10M-A, Thorlabs, USA) mounted on a computer controlled high precision rotational mount (PRM1/M-27E, Thorlabs, USA). The beam is then externally reflected from a 45°-90°-45° BK7 prism (PS912, Thorlabs, USA, refractive index n = 1.516) mounted on a precision rotation stage. The reflected beam is passed through another rotatable Glan-Thompson linear polarizer $P_2$, acting as the post selecting device. The resulting beam shift is detected either by using a quadrant photodiode (PDQ80A, Thorlabs, USA) or by a CCD camera (MP3.3-RTV-R-CLR-10-C, Singapore, 2048 × 1536 square pixels, pixel dimension 3.45 μm, 3 × 3 binning). In order to realize *scheme 1* of weak measurement, the polarization axis of $P_1$ is first set to +45° (with respect to the horizontal axis) so that it projects the state $|\psi in\rangle = [1, 1]^T$. As previously noted, the state of polarization of the reflected beam (pre selected state) is different from the input state but remains linear (Eq. 5), accordingly the analyzer ($P_2$) is rotated to minimize the intensity transmitted through it. At this position of $P_2$, the post selected state is exactly orthogonal to the pre selected state (Eq. 5). We then perform two measurements by changing the polarization axis of $P_2$ to $\pm\varepsilon$ away from this position. The corresponding shift in the centroid of the reflected beam between the two post selected states is recorded using either a QPD or a CCD (the large amplification of the beam shifts enabled detection using CCD also). The weak measurement *scheme 2* is realized in a similar way. The only difference is that here the polarization axis of $P_1$ is set to 0° or 90° in order to project the input p or s-polarization states. The beam shift measurements were performed as a function of angle of incidence ($\theta = 30°$ - $70°$), for varying small angle $\varepsilon$ ($\varepsilon$ =0.005 rad to 0.021 rad) and for varying propagation distance z. Coupling and simultaneous weak value amplification of the *angular* GH and IF shifts in the first scheme, manifested as a diagonal shift of the centroid of the reflected beam. The second scheme on the other hand, performs weak measurements of the IF shift alone, leading to a pure transversal shift of the beam's centroid, as demonstrated below.

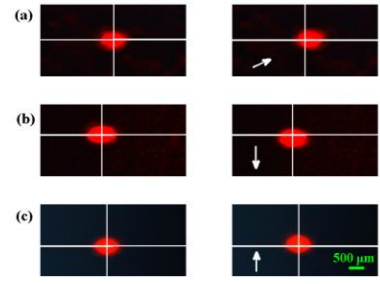

**Fig.2**. (a) Manifestation of coupling and simultaneous weak value amplification of the *angular* GH and the IF shifts. Diagonal shift in beam's centroid between the two post selected states +ε (left panel) and -ε (right panel) away from the orthogonal state for the input state $|\psi in\rangle = [1, 1]^T$ (scheme 1) is apparent. The observed transverse shift in beam's centroid for input states (b) $|\psi in\rangle = [1, 0]^T$ (scheme 2, p-state) and (c) $|\psi in\rangle = [0, 1]^T$ (scheme 2, s-state).

**Figure 2** illustrates the shifts in the centroid of the reflected beam as observed by employing the weak measurement schemes. These results are for the following experimental parameters: $\omega_0$ = 130μm, $\theta = 35°$, z = 46 cm, $\varepsilon$ = 0.007 rad. The amplified beam shifts for input states (a) $|\psi in\rangle = [1, 1]^T$ (scheme 1), (b) $|\psi in\rangle = [1, 0]^T$ (scheme 2, p-state) and (c) $|\psi in\rangle = [0, 1]^T$ (scheme 2, s-state) are displayed. Large diagonal shift in the beam's centroid between the two post selected states ($\pm \varepsilon$ of Eq. 5) is apparent for the input $[1, 1]^T$ state. The beam shifts for input p or s-polarization states (post selection according to Eq. 8), on the other hand, is entirely along the transverse direction. As expected, the shifts are in opposite direction for input p and s states since the amplitude reflection coefficients, $r_p$ and $r_s$, have opposite signs (out of phase by $\pi$) for this angle of incidence ($\theta < \theta_B$) and $|r_p| << |r_s|$.

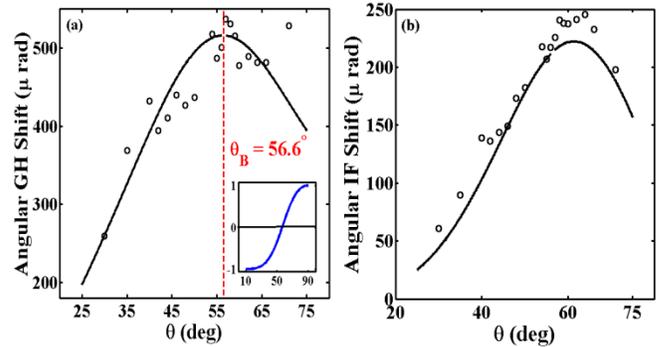

**Fig.3**. The dependence of (a) the angular GH shift and (b) the angular IF shift on the angle of incidence $\theta$ for the input $|\psi in\rangle = [1, 1]^T$ state (weak measurement scheme 1). Symbols represent experimental data and the lines are corresponding theoretical predictions (Eq. 7a and 7b respectively). The agreement *is* seen to be excellent on either side of the Brewster angle ($\theta_B \sim 56.6°$). The inset of 2a shows the computed angular dependence of the factor $2r_p r_s /(r_p^2+r_s^2)$.

In **Figure 3a and 3b**, we show the dependence of the angular GH and IF shifts (respectively) on the angle of incidence ($\theta$) for the input $|\psi in\rangle = [1, 1]^T$ state. The set of measurements reported here are for z =46 cm, $\varepsilon$ =0.007 rad. The magnitudes of the angular shifts were generated using the measured distances between the beam's centroid and the propagation distance z. Shifts in beam's centroid in a direction parallel ($<x>$) and perpendicular ($<y>$) to the plane of incidence were used separately for the determination of the angular GH and the IF shifts respectively. The corresponding theoretical predictions (using Eq. 7a and 7b) are also shown in the figures. Several observations are in place. The angular GH shift (Fig. 3a) exhibits strong dependence on $\theta$ and increases sharply as $\theta$ approaches the Brewster angle ($\theta \to \theta_B$). As shown previously in context to ordinary beam shift measurements, when $\theta$ is sufficiently close to $\theta_B$ [$(\theta - \theta_B) \leq 2\theta_o$, $\theta_o$ is the angular spread of the beam = $\lambda/\pi\omega_o$] incorporation of higher order corrections in the expression for the angular GH shift (Eq. 2) removes the singular behavior and leads to a dispersive resonance-like behavior of shift across $\theta_B$ [13,15,16]. We however, did not probe such behavior because the closest values for $\theta$ to the Brewster angle ($\theta$=55° and 57° approaching from the lower and the higher end respectively; $\theta_B$ ~ 56.6°) in our measurements were always away by $(\theta - \theta_B) > 2\theta_o$ ($\theta_o$ ~ 0.09°). Never-the-less, the observed magnitudes of the amplified angular GH shifts are in excellent agreement with the theoretical predictions (Eq. 7a), on either side of the Brewster angle ($\theta = 30° – 70°$ in the limit $(\theta - \theta_B) > 2\theta_o$). The angular IF shifts (Fig. 3b) also exhibit strong dependence on $\theta$. However, unlike angular GH shift, the IF shift peaks at an angle $\theta$ ~ 60° (>$\theta_B$). Once again, the magnitudes of the shift are in reasonable agreement with the theoretical predictions for all values of $\theta$ in the limit $(\theta - \theta_B) > 2\theta_o$. The relative signs of the shifts across the Brewster angle worth a brief mention. Note that in ordinary beam shift measurements, the angular GH shift exhibits reversal of sign (shift in opposite direction) across $\theta_B$ due to the change in sign of the reflection coefficient $r_p$ (phase change of $\pi$) [16]. However, the presence of the additional factor $2r_p r_s /(r_p^2+r_s^2)$ in this particular weak measurement scheme (Eq. 7) apparently removes this sign reversal (angular dependence of this factor for air/glass interface is shown in the inset of Fig. 3a for convenience).

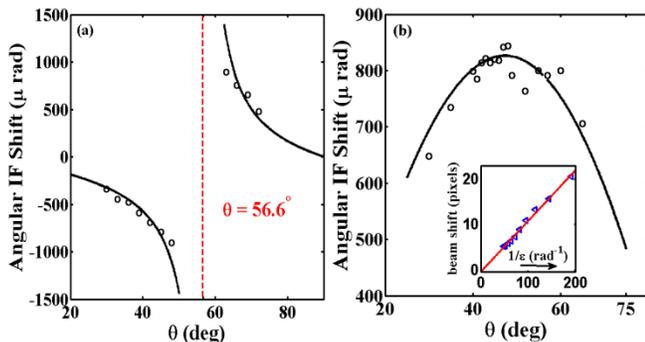

**Fig.4**. The dependence of the angular IF shift on $\theta$ for input (a) p linear polarization state and (b) s linear polarization state (scheme 2). Symbols represent experimental data and the lines are theoretical predictions (Eq. 9). The inset of 4b shows the weak value amplification of beam shift as a function of ε. The linear scaling of the beam shifts with 1/ε demonstrates reliable amplification.

**Figure 4a and 4b** summarize the results of weak measurements using scheme 2 for the amplification of IF shift alone. The observed trends can be summarized as –

(i) In agreement with Eq. 9, the IF shifts for both p (Fig. 4a) and s (Fig. 4b) polarization states exhibit considerable variation with $\theta$ and the shifts increase sharply as $\theta \to \theta_B$. (b) While the IF shift exhibits sign reversal across Brewster angle for input p polarization state, no such reversal of shift is observed for input s polarization state. As apparent from Eq. 9, this arises because the ratio $r_s/r_p$ is negative (positive) for $\theta < \theta_B$ ($\theta > \theta_B$) and that for the measured angular range ($\theta = 30° - 70°$),|$r_p$| << |$r_s$|. Finally, the amplified angular GH and IF shifts in either of the weak measurement schemes exhibited the expected 1/ε dependence. Representative results for amplified IF shift as a function of ε is shown in the inset of Figure 4b. Linear scaling of the shifts with 1/ε and the observed excellent agreement between theoretical predictions and measurements provide concrete evidence of reliable amplification of both the angular GH and IF shifts in partial reflection using weak measurement schemes.

To summarize, we have demonstrated weak value amplification of both *angular* GH and IF shifts in *partial reflection*. Among the two weak measurement schemes employed for this purpose, the first one enabled simultaneous amplification of both the angular GH and IF shifts, the second one on the other hand, amplified angular IF shift alone by nullifying the weak measurement on angular GH shift. Weak measurements employing either of these schemes faithfully amplified the beam shifts in partial reflection for any angle of incidence. The demonstrated novel ability to amplify the tiny *angular* beam shifts particularly in case of *partial reflection* from dielectric interfaces may prove to be valuable for developing ultra sensitive sensors and nano probes.

**References**
1. F. Goos and H. Hänchen, Annals of Physics, **1**, 333 (1947).
2. C. Imbert, Physical Review D, **5**, 787 (1972).
3. K.Y. Bliokh, A Aiello, Journal of Optics, **15**, 014001 (2013).
4. A. Aiello, New. J. Phys., **14**, 013058 (2012).
5. O. Hosten and P. Kwiat, Science, **319**, 787 (2008).
6. K. Y. Bliokh, A. Niv, V. Kleiner and E. Hasman, Nat. Photon., **2**, 748 (2008).
7. J.B. Gotte, M.R. Dennis, New. J. Phys., **14**, 073016 (2012).
8. M. R. Dennis and J. B. Gotte, New J. Phys., **14**, 073013 (2012).
9. F. Toppel, M. Ornigotti, and A. Aiello, New. J. Phys., **15**, 113059 (2013).
10. Y. Gorodetski, K. Y. Bliokh, B. Stein, C. Genet, N. Shitrit, V. Kleiner, E. Hasman, and T. W. Ebbesen, Phys. Rev. Lett., **109**, 013901 (2012).
11. G. Jayaswal, G. Mistura and M. Merano, Opt. Lett., **38**, 1232 (2013).
12. G. Jayaswal, G. Mistura and M. Merano, Opt. Lett., **39**, 2266 (2014).
13. J.B. Gotte, M.R. Dennis, Opt. Lett., **38**, 2295 (2013).
14. A. Aiello and J. P. Woerdman, Opt. Lett., **33**, 1437 (2008).
15. A. Aiello A and J. P. Woerdman, arXiv:0903.3730 (2009).
16. M. Merano, A. Aiello, M.P. van Exter and J. P. Woerdman, Nat. Photon., **3**, 337 (2009).